\documentclass[iop]{emulateapj}
\submitted{Received 2011 November 1; accepted 2011 December 13}
\journalinfo{The Astrophysical Journal Letters}
\usepackage{apjfonts}
\shorttitle{RAPID CHANGES OF PHOTOSPHERIC MAGNETIC FIELD}
\shortauthors{LIU ET AL.}
\usepackage{hyperref,url}
\usepackage{breakurl}
\def\mathbi#1{\textbf{\em #1}}
\newcommand{\hinode}{\textit{Hinode}}
\newcommand{\Hsi}{\textit{Reuven Ramaty High Energy Solar Spectroscopic Imager}}
\newcommand{\hsi}{\textit{RHESSI}}
\newcommand{\goes}{\textit{GOES}}
\newcommand{\sm}{$\sim$}
\newcommand{\Sdo}{\textit{Solar Dynamics Observatory}}
\newcommand{\sdo}{\textit{SDO}}
\newcommand{\hmi}{Helioseismic and Magnetic Imager}
\newcommand{\aia}{Atmospheric Imaging Assembly}

\begin{document}

\title{RAPID CHANGES OF PHOTOSPHERIC MAGNETIC FIELD AFTER TETHER-CUTTING RECONNECTION AND MAGNETIC IMPLOSION}
\author{Chang Liu\altaffilmark{1}, Na Deng\altaffilmark{1,2}, Rui Liu\altaffilmark{1}, Jeongwoo Lee\altaffilmark{3}, Thomas Wiegelmann\altaffilmark{4}, Ju Jing\altaffilmark{1}, Yan Xu\altaffilmark{1}, Shuo Wang\altaffilmark{1},\\and Haimin Wang\altaffilmark{1}}
\affil{1. Space Weather Research Laboratory, New Jersey Institute of Technology, University Heights, Newark, NJ 07102-1982, USA; chang.liu@njit.edu}
\affil{2. Department of Physics and Astronomy, California State University, Northridge, CA 91330-8268, USA}
\affil{3. Physics Department, New Jersey Institute of Technology, University Heights, Newark, NJ 07102-1982, USA}
\affil{4. Max Planck Institut f{\"u}r Sonnensystemforschung (MPS), Max-Planck-Strasse 2, 37191 Katlenburg-Lindau, Germany}

\begin{abstract}
The rapid, irreversible change of the photospheric magnetic field has been recognized as an important element of the solar flare process. This Letter reports such a rapid change of magnetic fields during the 2011 February 13 M6.6 flare in NOAA AR 11158 that we found from the vector magnetograms of the Helioseismic and Magnetic Imager with 12-min cadence. High-resolution magnetograms of Hinode that are available at \sm$-$5.5, $-$1.5, 1.5, and 4 hrs relative to the flare maximum are used to reconstruct three-dimensional coronal magnetic field under the nonlinear force-free field (NLFFF) assumption. UV and hard X-ray images are also used to illuminate the magnetic field evolution and energy release. The rapid change is mainly detected by HMI in a compact region lying in the center of the magnetic sigmoid, where the mean horizontal field strength exhibited a significant increase by 28\%. The region lies between the initial strong UV and hard X-ray sources in the chromosphere, which are cospatial with the central feet of the sigmoid according to the NLFFF model. The NLFFF model further shows that strong coronal currents are concentrated immediately above the region, and that more intriguingly, the coronal current system underwent an apparent downward collapse after the sigmoid eruption. These results are discussed in favor of both the tether-cutting reconnection producing the flare and the ensuing implosion of the coronal field resulting from the energy release.
\end{abstract}


\section{INTRODUCTION}
It has been known that the long-term evolution of photospheric magnetic field (PMF) driven by new flux emergence and surface flows plays important roles in building up free energy in the corona, and that this free magnetic energy powers flares and coronal mass ejections \citep[CMEs;][]{priest02}. On the other hand, a short-term variation of the PMF associated with flares has not been considered because PMFs are strongly line-tied to the dense high-$\beta$ photosphere and thus are thought unlikely to be altered by any flare-related  disturbances created in the tenuous low-$\beta$ corona. Only recently, a back reaction of the coronal magnetic field (CMF) on the PMF during the reconfiguration of the CMF has been seriously considered from the theoretical point of view. The idea is that the CMF should contract inward, as the magnetic energy of the CMF decreases after flares and/or CMEs \citep[called ``implosion'';][]{hudson00}. This may make the PMF be oriented more horizontally, resulting in a Lorentz force acting downwardly on the solar surface \citep{fisher10,hudson08}. In fact, a few theoretical flare models predict the appearance of more horizontal PMFs after flares/CMEs in the form of newly formed, low-lying field lines close to the surface \citep[e.g.,][]{melrose97,moore01}.

The phenomenon of coronal implosion has been reported in several recent observational studies \citep{liur+implosion09,liur+wang09,liur+wang10}. The related phenomenon of the rapid change of the PMF to incline more horizontally after flares had also been reported almost two decades ago \citep{wang92,wang94}. Namely, the transverse field near the magnetic polarity inversion line (PIL) at the core flaring region is enhanced rapidly and irreversibly, which is often accompanied by an increase of magnetic shear. Such a rapid change of vector PMF change closely associated with flares/CMEs has been consistently found later on using ground-based \citep{schmieder94,wang02b,wang04,wang+liu05,liu05,wang07a,wang10} and \hinode\ observations \citep{jing08,li09}. Indirect evidence includes the unbalanced flux evolution of the line-of-sight magnetic field \citep{wang10}, the variation of the sunspot white-light structure in flaring regions \citep{wang04a,deng05,liu05,chen07b}, and the change in the pattern of the penumbral Evershed flow \citep{deng11}. Additional support also comes from the three-dimensional (3D) MHD simulations of an emerging twisted flux tube \citep{fan10}, the eruption of which induces similar impact on the low-altitude magnetic fields \citep{li10}.

\begin{figure*}[t]
\epsscale{1}
\plotone{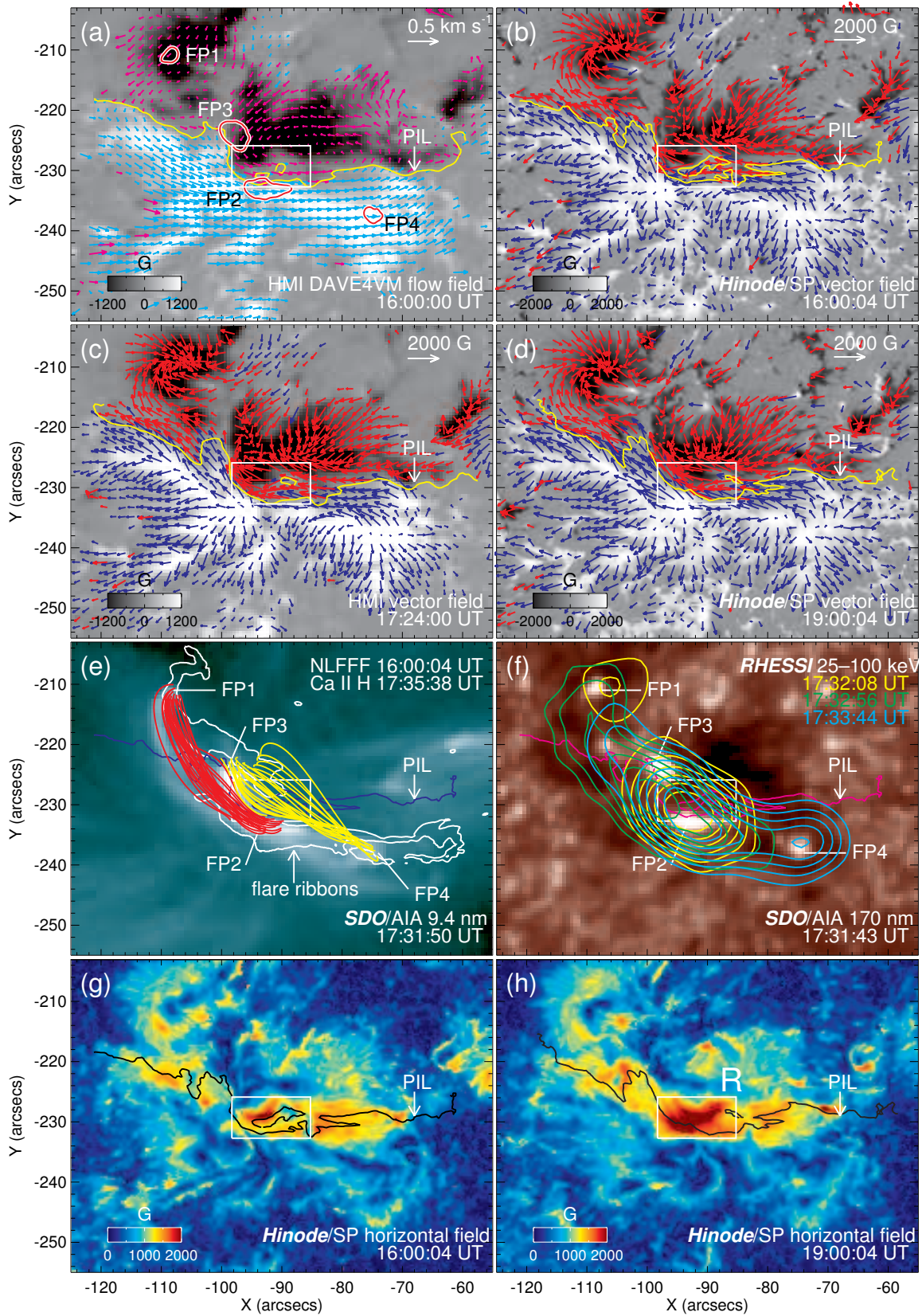}
\caption{$B_v$ maps are superimposed with arrows (for clarity, those in negative and positive fields are coded in different colors) representing horizontal plasma flows (a) and magnetic field vectors (b--d). The NLFFF lines in (e) stem from the four flare kernels FP1--FP4 in 1700~\AA\ (f) at the flare onset. The white contours in (e) outline the double J-shaped flare ribbons in Ca~{\sc ii}~H. Contours (with levels of 38\%, 48\%, 58\%, 68\%, 78\%, 88\% of the maximum flux) in (f) denote \hsi\ CLEAN images reconstructed with detectors 2--8 (except that 17:32:08~UT image is made without the detector 2 due to the relatively low counts at this time). The white box overplotted on all panels denotes the region R under study with enhanced $B_h$ after the flare. \label{f1}}
\end{figure*}

While the evolution of vector PMFs closely associated with flares were mostly studied using ground-based observations subject to seeing variations, \citet{wang+shuo11} and \citet{sun11} used data from the \hmi~\cite[HMI;][]{schou11} on board the recently launched \Sdo\ (\sdo) to find a rapid enhancement of transverse field associated with the 2011 February 15 X2.2 flare. Since solar activity is now increasing, this is an opportune moment to advance the study of the flare-related evolution of the PMF using the state-of-the-art, seeing-free photospheric vector magnetograms acquired by space missions.

\begin{figure*}[t]
\epsscale{.8}
\plotone{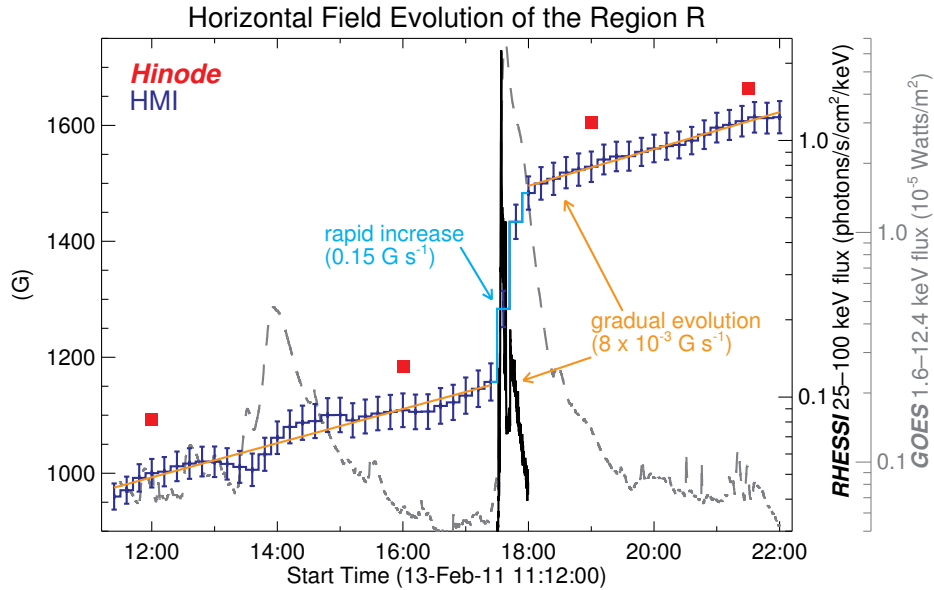}
\caption{Temporal evolution of $\langle B_h \rangle$ of the region R in Figure~\ref{f1}, in comparison with the flare light curves of soft and hard X-rays. The mean error of $B_h$ of the region R at each time instance measured by HMI is plotted as error bars. The error estimates for \hinode\ data are currently not usable$^2$. \label{f2}}
\end{figure*}

In this Letter, we investigate the PMF evolution associated with an M6.6 flare on 2011 February 13 that was well covered by both HMI and \hinode\ (see Section~\ref{data}), in which the transverse field enhancement at the flaring PIL is unequivocally detected. We will discuss the rapid change of the 3D CMF associated with that of the PMF using nonlinear force-free field (NLFFF) extrapolations from the active region.

\section{OBSERVATIONS AND DATA REDUCTION} \label{data}
The HMI instrument obtains filtergrams in six polarization states at six wavelengths along the Fe~{\sc i} 6173~\AA\ spectral line to compute Stokes parameters {\it IQUV}. The inverted and disambiguated vector magnetic field data for the NOAA AR 11158 with a 12~minute cadence and 0.5\arcsec\ pixel size were recently released by the HMI team\footnote{\url{http://jsoc.stanford.edu/jsocwiki/VectorPaper}} \citep{hoeksema11}. We used the vector magnetograms remapped to heliographic coordinates to (1) investigate the rapid PMF evolution associated with the M6.6 flare on 2011 February 13, which started at 17:28~UT, peaked at 17:38~UT, and ended at 17:47~UT in terms of \goes\ 1--8~\AA\ flux, and (2) trace the surface flows with the differential affine velocity estimator for vector magnetograms \citep[DAVE4VM;][]{schuck08}.

The Spectro-Polarimeter (SP) of the Solar Optical Telescope \citep[SOT;][]{tsuneta08} on board \hinode\ obtains polarization profiles of two magnetically sensitive Fe lines at 630.15 and 630.25~nm to generate Stokes images, which are then reduced with the \hinode/SP calibration/inversion pipeline to retrieve the vector PMF using the Milne-Eddington gRid Linear Inversion Network (MERLIN)\footnote{\url{http://sot.lmsal.com/data/sot/level2dd/sotsp_level2_description.html}}. \hinode/SP scanned the nearly entire isolated, near disk-center (about S20$^{\circ}$, E04$^{\circ}$) AR 11158 at four time bins (12:00--12:32, 16:00--16:32, 19:00--19:32, and 21:30--22:02~UT) close to the M6.6 flare. We further processed the resulted vector magnetograms with a resolution of 0$''$.32~pixel$^{-1}$ to (1) resolve the 180$^{\circ}$ azimuthal ambiguity using the AZAM code \citep{leka09b} based on the ``minimum energy'' approach \citep{metcalf06}, and (2) remove the projection effects by transforming the observed fields to heliographic coordinates \citep{gary_hagyard90}. Taking advantage of the high spatial resolution and high polarization accuracy of \hinode, we examine the CMF evolution by constructing NLFFF models using the ``weighted optimization'' method \citep{wiegelmann04} after preprocessing the photospheric boundary to best suit the force-free condition \citep{wiegelmann06}. The calculation was performed using 2~$\times$~2 rebinned magnetograms within a box of 236~$\times$~256~$\times$~256 uniform grid points, which corresponds to \sm110~$\times$~120~$\times$~120~Mm$^3$ of balanced magnetic fluxes (with a ratio of opposite surface fluxes of \sm99\%--102\%).

The temporal and spatial relationship between the PMF change and flare energy release can provide important clues concerning the eruption mechanism. The evolution of the flare hard X-ray (HXR) emission was entirely registered by the \Hsi\ \citep[\hsi;][]{lin02}. CLEAN images \citep{hurford02} in the nonthermal energy range (25--100~keV) showing the flare footpoints were reconstructed using the front segments of detectors 2--8 with \sm48~s integration time throughout the event. To provide the observational context from the low chromosphere to the coronal, we also used 1700~\AA\ continuum (5000~K) and 94~\AA\ (Fe~{\sc xviii}; 6~MK) images taken by the \aia\ \citep[AIA;][]{leman11} on board \sdo, and Ca~{\sc ii}~H images taken by \hinode/SOT.

\section{Flare-related PMF Changes}
The M6.6 flare is initiated at the center of the active region, where opposite magnetic flux concentrations undergo an overall counterclockwise, rotation-like motion as clearly tracked by DAVE4VM (Figure~\ref{f1}(a)), resulting in highly sheared fields along the PIL (Figures~\ref{f1}(b) and (c)). A compact region R at the central PIL (the boxed region in Figure~\ref{f1}) is readily identified in the HMI vector magnetograms to have a pronounced, \sm28\% enhancement of the mean horizontal field strength from $\langle B_h \rangle \simeq$1160~G at 17:24~UT immediately before the flare to $\langle B_h \rangle \simeq$1480~G at 18:00~UT right after the flare (see Figure~\ref{f2}) in \sm30~minutes, with a rate of \sm0.15~G~s$^{-1}$. The change-over time is cotemporal with the rapid rising of soft X-ray flux and peaking of HXR emissions. Standing out over the long-term evolution of $\langle B_h \rangle$, which increases much more slowly with time (at \sm8~$\times 10^{-3}$~G~s$^{-1}$), this change occurs in a stepwise fashion, coincident with the occurrence of the M6.6 flare, and the enhanced $\langle B_h \rangle$ survives the flare and persists permanently. Consistent with our previous results \citep[][and references therein]{wang10}, this rapid PMF change also leads to a more horizontal state of the PMF in the region R after the flare.

\begin{figure}[t]
\epsscale{1.17}
\plotone{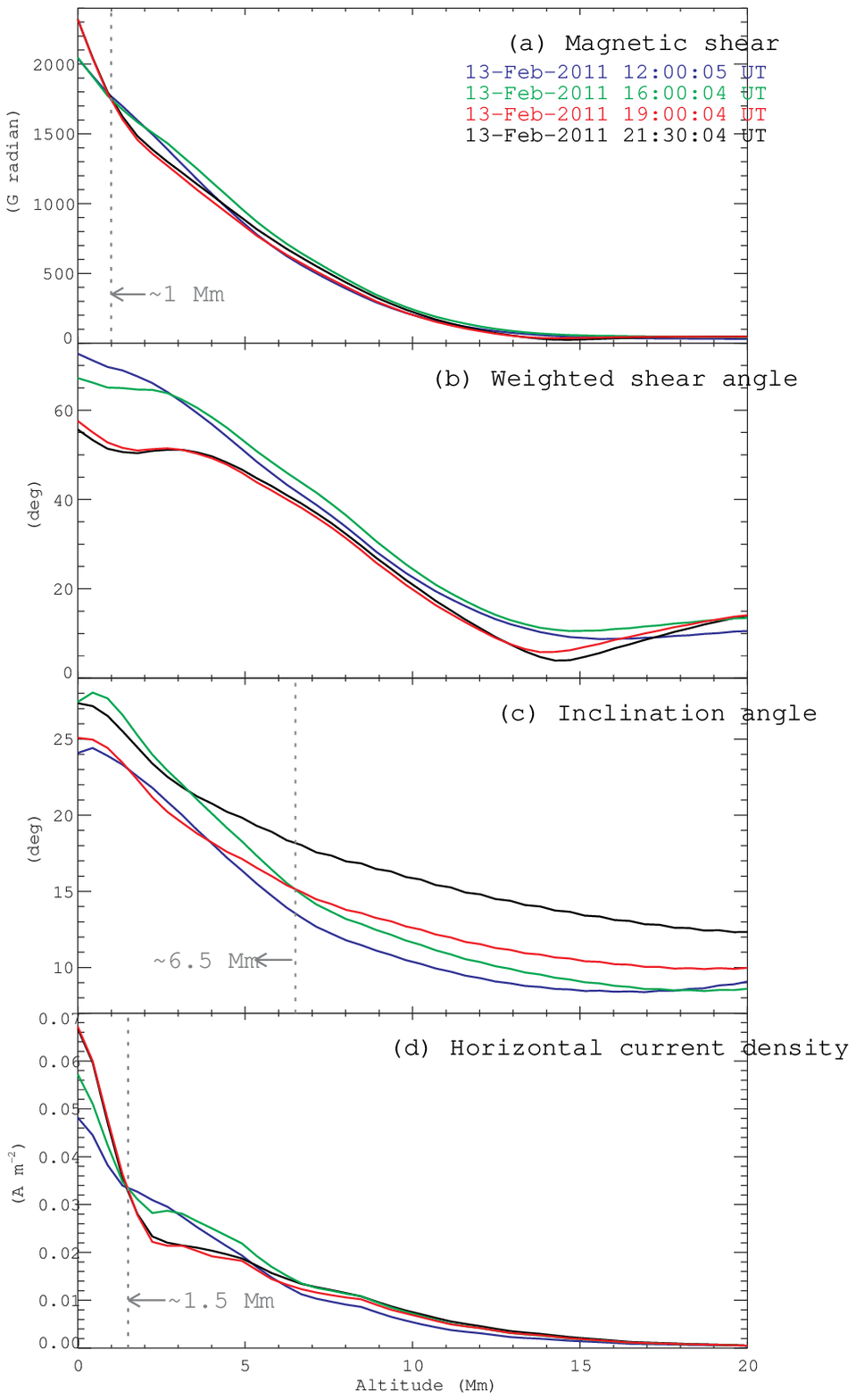}
\caption{$\langle \tilde{S} \rangle$, $\mathring{S}$, $\langle \varphi \rangle$, and $\langle J_h \rangle$ of the region R in Figure~\ref{f1} as a function of altitude with a step size of \sm0.46~Mm, at four times of \hinode\ scans. \label{f3}}
\end{figure}

Similar to the HMI result, the $\langle B_h \rangle$ within the region R is observed by Hinode to increase from $\simeq$1090~G (12~UT) and 1180~G (16~UT) before the flare to $\simeq$1600~G (19~UT) and 1660~G (21:30~UT) after the flare (cf. Figures~\ref{f1}(b) and (d), and (g) and (h); also see Figure~\ref{f2}). It is particularly interesting that there are four compact flare kernels FP1--FP4 in 1700~\AA\ at the flare onset, and that the region R is sandwiched between the much stronger FP2 and FP3 (Figure~\ref{f1}(f)). Following \citet{liu10}, we use the locations of the flare kernels together with a preflare NLFFF model at 16:00~UT based on the \hinode\ data to shed light on the configuration of the flaring magnetic fields. The extrapolated field lines connecting among FP1--FP4 as plotted in Figure~\ref{f1}(e) apparently form a S-shaped sigmoidal structure with right-bearing loops connecting FP1--FP2 and FP3--FP4, which is similar to the overall shape of coronal loops in EUVs. This is consistent with the calculation that the magnetic helicity in this active region has a positive sign \citep{liuy11}. It is furthermore noticeable that (1) the early HXR emission before \sm17:32:30~UT is imaged as footpoint-like sources cospatial with FP1 and FP2 (Figure~\ref{f1}(f)), and (2) ribbon-like HXR emissions mimicking flaring loops that connect among FP1--FP4 dominate later on, when flare kernels evolve to double J-shaped ribbons (the white contours in Figure~\ref{f1}(e)). These lead us to a picture in which the flare could be triggered by the tether-cutting reconnection \citep{moore95,moore01,xu10} between the two sets of sigmoidal loops FP1--FP2 and FP3--FP4. The reconnected large-scale fields FP1--FP4 are seen to erupt outward to become the CME, and the newly formed smaller loops FP2--FP3 lying close to the surface could then account for the detected enhancement of $B_h$ at the region R. We also note that (1) the aforementioned tether-cutting reconnection could be driven by the converging flows from FP3 toward FP2 \citep[see Figure~\ref{f1}(a);][]{moore01,liur10}, and (2) the region R at the PIL lies between the flare kernels/ribbons (see Figures~\ref{f1}(e) and (f)), hence the observed $B_h$ enhancement is unlikely to be affected by flare emissions \citep{patterson81,qiu03}.

\section{Change of CMF Topology Inferred from NLFFF Model} \label{nlfff}
To quantify the flare-induced change of the CMF in 3D, we first define several characteristic quantities of magnetic field. The nonpotentiality can be evaluated using magnetic shear $\tilde{S}$, which is computed here as the product of field strength and shear angle \citep{wang94,wang06shear,jing08}:

\begin{equation}
\tilde{S}=B \cdot \theta \ ,
\end{equation}

\noindent where $B=|\mathbi{B}|$, $\theta={\rm cos}^{-1}(\mathbi{B} \cdot \mathbi{B}_p)/(BB_p)$, and the subscript $p$ represents the potential field derived using the Green's function method \citep{metcalf08}. The weighted shear angle over a region of interest with $n$ pixels is then

\begin{equation}
\mathring{S}=\frac{\sum_{i} \tilde{S}_i}{\sum_i B_i} \ , \ {\rm where} \ i=1,2,...,n \ .
\end{equation}
	
\noindent In this study, the magnetic inclination angle $\varphi$ is defined as the azimuthal difference relative to the horizontal plane:

\begin{equation}
\varphi = {\rm tan}^{-1} \frac{|B_v|}{(B_x^2+B_y^2)^{1/2}} \ .
\end{equation}

\noindent Following \citet{hudson08} and \citet{fisher10}, the change of Lorentz force acting on the surface in the vertical direction given a change of PMF $\delta \mathbi{B}$ can be quantified as

\begin{equation}
\delta F_{v \rm , \ downward} =\frac{1}{4 \pi} \int dA (-B_x\delta B_x-B_y\delta B_y + B_v\delta B_v) \ .
\end{equation}

\begin{figure*}[t]
\epsscale{1.05}
\plotone{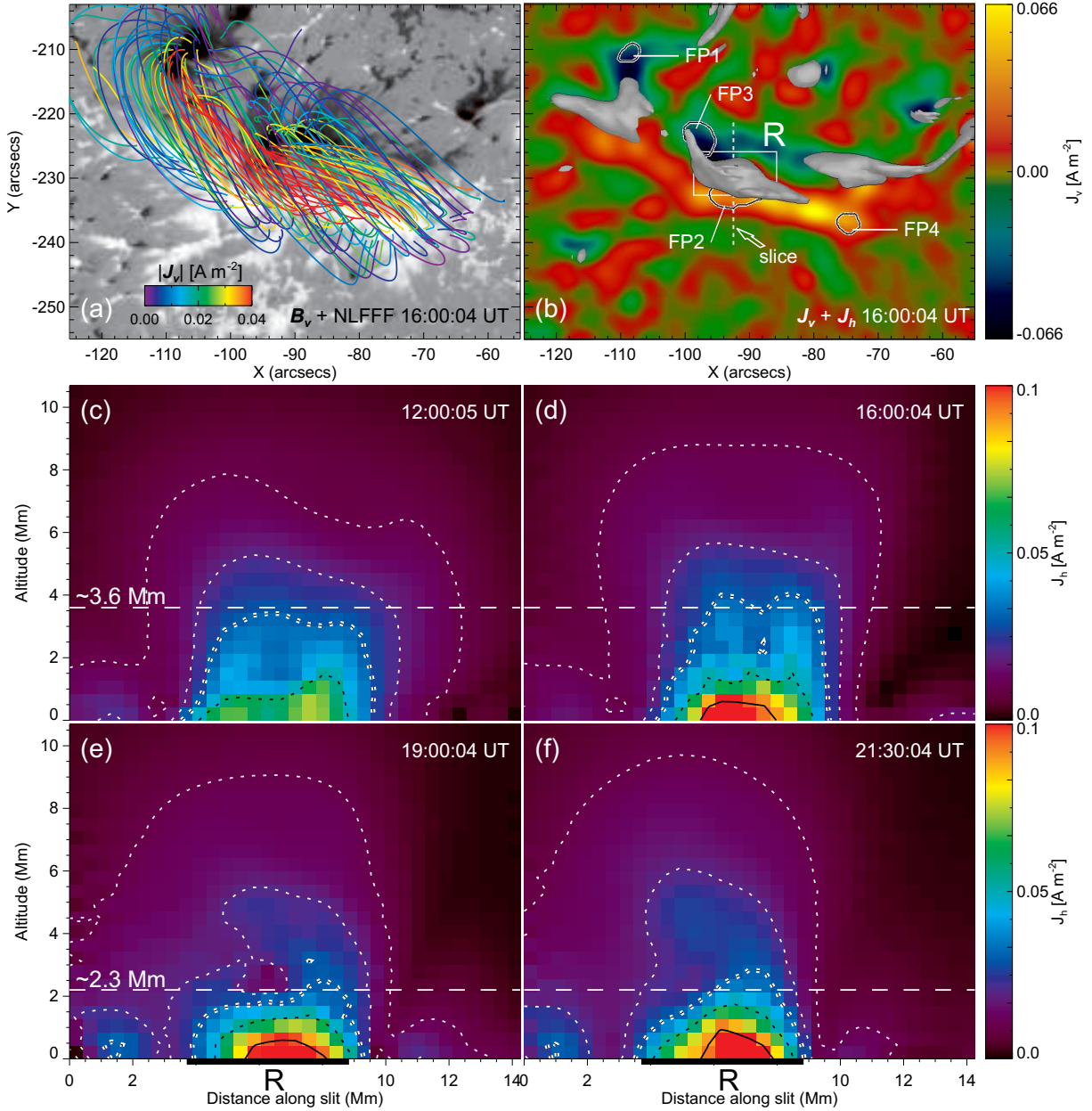}
\caption{(a) A preflare \hinode\ $B_v$ map superimposed with the NLFFF lines colored according to $|J_v|$ at footpoints. (b) A preflare $J_v$ map superimposed with isosurfaces of $J_h=0.05$~A~m$^{-2}$ (gray). (c--f) Pre- and postflare distribution of $J_h$ in a vertical cross section with its bottom side denoted in (b) as the dotted line. The contour levels are 0.01, 0.02, 0.03, 0.05, and 0.09~A~m$^{-2}$. \label{f4}}
\end{figure*}

We calculate $\langle \tilde{S} \rangle$, $\mathring{S}$, and $\langle \varphi \rangle$ of the region R as a function of altitude, and make the comparison among the four NLFFF extrapolations using \hinode\ scans in Figures~\ref{f3}(a)--(c). It is clearly shown that (1) after the flare, $\langle \tilde{S} \rangle$ becomes larger from the surface up to \sm1~Mm but smaller above that height; meanwhile, $\mathring{S}$ decreases at nearly all heights. These suggest that while the CMF is relaxed in general following the flare energy release, the near-surface fields become more stressed due to a concentration of currents immediately above the region R (see below) with the new connectivity between FP2 and FP3 as a result of the tether-cutting reconnection. Similarly, both $\langle \tilde{S} \rangle$ and $\mathring{S}$ enhance near the surface in the event of \citet{jing08}, which was, however, attributed to the newly emerged sheared fields after the flare. (2) the whole $\langle \varphi \rangle$ profile shows increase at nearly all heights in the preflare and postflare states presumably due to the long-term evolution driven by new flux emergence. Noticeably, below \sm6.5~Mm $\langle \varphi \rangle$ decreases across the flare time with larger drop nearer the surface. This clearly indicates that the magnetic field of and at low altitudes above the region R turns more inclined toward the surface, which agrees with the implication from the $\langle B_h \rangle$ enhancement at the surface and may evidence the coronal implosion scenario \citep{hudson00} in which the CMF collapses downward after releasing energy. The implosion may result in a downward impulse due to the change of Lorentz force $\delta F_z$ acting on the region R, which is estimated to be $-2 \times 10^{22}$~dynes.

In order to visualize the collapse of the CMF using the NLFFF model, we further derive the total electric current density and its horizontal component $|J_h|=(J_x^2+J_y^2)^{1/2}$. We note the followings based on the results presented in Figure~\ref{f4}.

First, associated with the sigmoidal fields at the flare core region (highly sheared field shown in the red-yellow color with overarching higher fields), strong currents flow along the sheared flaring PIL. The channel-like portion of the highest density $J_h \gtrsim 0.05$~A~m$^{-2}$ is located directly above the region R (with height $\lesssim$1.5~Mm; Figures~\ref{f4}(b)--(d)). This small and low-lying volume with the concentration of current is presumably the site of the tether-cutting reconnection \citep{moore01}. It is worthwhile to point out that the most intense, initial flare kernels FP2 and FP3 as well as the early HXR emission centered at FP2 are not cospatial with the maximum vertical electric current density $J_v$ but instead positioned on the two sides next to the above intense horizontal current channel (Figures~\ref{f1}(f) and \ref{f4}(b)). This suggests that it is the instability of the horizontal current that is responsible for electron acceleration during the early phase of the flare \citep{leka93}.

\begin{figure}[t]
\epsscale{1.15}
\plotone{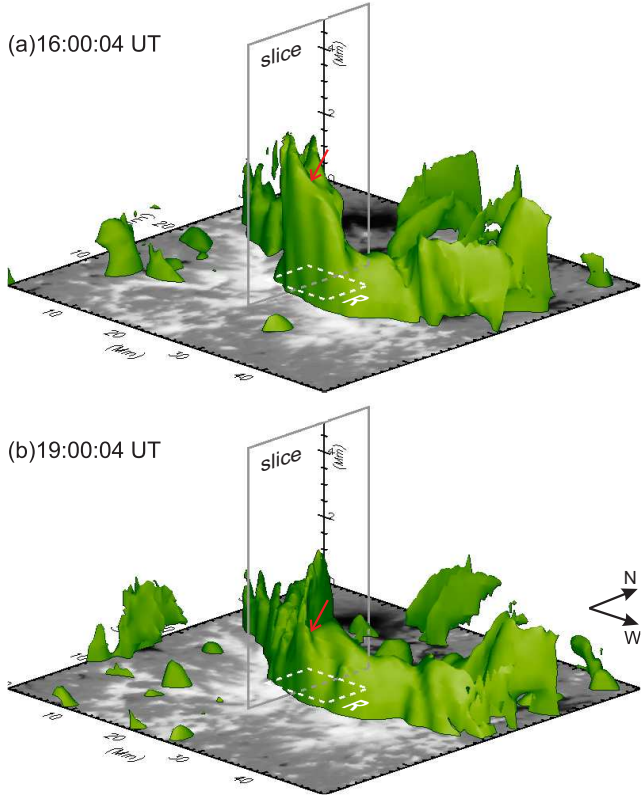}
\caption{Isosurfaces of $J_h=0.03$~A~m$^{-2}$ viewed from 30$^{\circ}$ relative to the horizontal plane and 45$^{\circ}$ clockwise about the vertical direction. The slice is the same as used in Figure~\ref{f4}. \label{f5}}
\end{figure}

Second, we cut a slice through the 3D CMF (with its surface intersection denoted in Figure~\ref{f4}(b) as the dotted line) and plot the distribution of $J_h$ in this cross section for the pre- (Figures~\ref{f4}(c) and (d)) and postflare (Figures~\ref{f4}(e) and (f)) states. The downward collapse of the CMF after the flare is clearly demonstrated by the evolution of $J_h$, which is most discernible as the contour level of 0.03~A~m$^{-2}$ (white bordered, dotted lines) decreases from \sm3.6 to 2.3~Mm above the region R. Comparison between the pre- and postflare 3D isosurfaces of $J_h$ (Figure~\ref{f5}) also evidently indicates that the collapse predominantly occurs above the region R (pointed to by the red arrow) where the energy is expected to release as above mentioned. We further plot $\langle J_h \rangle$ in the region R as a function of altitude in Figure~\ref{f3}(d), in which it can be seen that (1) $\langle J_h \rangle$ becomes stronger at nearly all heights from 12 to 16~UT, which could be due to the long-term evolution of the active region (e.g., Figures~\ref{f2}), (2) from 16 to 19~UT, $\langle J_h \rangle$ further enhances below \sm1.5~Mm with larger increase nearer the surface while it tends to diminish above that height, and (3) $\langle J_h \rangle$ does not evolve significantly from 19:00 to 21:30~UT. Importantly, the enhancement of $\langle J_h \rangle$ across the flare duration is consistent with that of $\langle \tilde{S} \rangle$ at low altitudes, and is also in line with the model of \citet{melrose97}, in which smaller current-carrying loop is produced by the tether-cutting-like reconnection between the initial larger loops so that the current path moves downward closer to the surface. We tentatively suggest that the tether-cutting reconnection induces the CMF reconnection, which is accompanied by the ensuing collapse of the CMF due to energy release.

\section{SUMMARY AND DISCUSSION}
We have used the vector magnetograms from HMI and \hinode\ to analyze the change of the PMF associated with the 2011 February 13 M6.6 flare, and interpreted the results with the aid of NLFFF modeling based on the \hinode\ data under the context of 3D magnetic reconnection. Major results are summarized as follows.

\begin{enumerate}
\item A compact region R at the central flaring PIL shows a rapid and permanent enhancement of $\langle B_h \rangle$ by \sm320~G (\sm28\% of the preflare magnitude as observed by HMI) closely associated with the flare. A similar trend of $\langle B_h \rangle$ evolution is also detected by \hinode. Overall, the PMF at the region R becomes more inclined to the surface after the flare.

\item The magnetic fields lying immediately above R, which correspond to the central part of the sigmoidal region, are highly sheared and have strong horizontal electric currents. The most intense flare kernels are initially located at the two ends of this channel-like current system, implying that the start of electron acceleration could be closely associated with the strong horizontal currents \citep{leka93}. The configuration of the NLFFF lines is favorable for a subsequent tether-cutting reconnection \citep{moore01}, when disturbed by converging surface flows. This ends up with the short and low-lying loops, which explains the enhanced $B_h$ and $\tilde{S}$ at the region R.

\item Across the flare energy-release time, the nonpotentiality as represented by $\tilde{S}$ increases in the height range from the surface up to \sm1~Mm and then decreases at higher altitudes, which is closely correlated with the evolution of $J_h$. This is in accordance with the reconnection of current-carrying loops modeled by \citet{melrose97}. Remarkably, the vertical profiles of $J_h$ as well as $\varphi$ before and after the flare clearly demonstrate a downward collapse of the CMF by \sm1.3~Mm toward the surface, which is probably related to coronal implosion \citep{hudson00}.
\end{enumerate}

In summary, the change of the PMF and the corresponding change of the CMF observed in this study are not caused by the long-term development of the active region but occurred as rapid as the flare. The magnetic field change may involve two closely related physical processes: the tether-cut reconnection producing the flare and the ensuing collapse of the CMF resulting from the energy release.

\acknowledgments
\hinode\ is a Japanese mission developed and launched by ISAS/JAXA, with NAOJ as domestic partner and NASA and STFC (UK) as international partners. It is operated by these agencies in co-operation with ESA and NSC (Norway). \sdo\ is a mission for NASA's Living With a Star program. \hsi\ is a NASA Small Explorer. We thank the \sdo/HMI team for the free access to the newly released HMI vector magnetograms, which were used in revising the paper. We thank the referee for helpful comments. This work uses the DAVE4VM code written and developed by the Naval Research Laboratory. C.L., R.L., J.J., Y.X., S.W. and H.W. were supported by NSF grants AGS 08-19662, AGS 08-49453, AGS 09-36665, and AGS 08-39216, and NASA grants NNX 08AQ90G, NNX 08AJ23G, NNX 11AC05G, and NNX 11AQ55G. N.D. was supported by NASA grant NNX 08AQ32G. J.L. was supported by NSF grant AST 09-08344. T.W. was supported by DLR-grant 50OC0501.


\begin{thebibliography}{50}
\expandafter\ifx\csname natexlab\endcsname\relax\def\natexlab#1{#1}\fi

\bibitem[{{Chen} {et~al.}(2007){Chen}, {Liu}, {Song}, {Deng}, {Tan}, \&
  {Wang}}]{chen07b}
{Chen}, W., {Liu}, C., {Song}, H., {Deng}, N., {Tan}, C., \& {Wang}, H. 2007,
  Chin. J. Astron. Astrophys., 7, 733

\bibitem[{{Deng} {et~al.}(2011){Deng}, {Liu}, {Prasad Choudhary}, \&
  {Wang}}]{deng11}
{Deng}, N., {Liu}, C., {Prasad Choudhary}, D., \& {Wang}, H. 2011, \apjl, 733,
  L14+

\bibitem[{{Deng} {et~al.}(2005){Deng}, {Liu}, {Yang}, {Wang}, \&
  {Denker}}]{deng05}
{Deng}, N., {Liu}, C., {Yang}, G., {Wang}, H., \& {Denker}, C. 2005, \apj, 623,
  1195

\bibitem[{{Fan}(2010)}]{fan10}
{Fan}, Y. 2010, \apj, 719, 728

\bibitem[{{Fisher} {et~al.}(2010){Fisher}, {Bercik}, {Welsch}, \&
  {Hudson}}]{fisher10}
{Fisher}, G.~H., {Bercik}, D.~J., {Welsch}, B.~T., \& {Hudson}, H.~S. 2010,
  ArXiv e-prints, 1006.5247

\bibitem[{{Gary} \& {Hagyard}(1990)}]{gary_hagyard90}
{Gary}, G.~A., \& {Hagyard}, M.~J. 1990, \solphys, 126, 21

\bibitem[{{Hoeksema} {et~al.}(2011)}]{hoeksema11}
{Hoeksema}, J.~T., {et~al.} 2011, \solphys, to be submitted

\bibitem[{{Hudson}(2000)}]{hudson00}
{Hudson}, H.~S. 2000, \apjl, 531, L75

\bibitem[{{Hudson} {et~al.}(2008){Hudson}, {Fisher}, \& {Welsch}}]{hudson08}
{Hudson}, H.~S., {Fisher}, G.~H., \& {Welsch}, B.~T. 2008, in Astronomical
  Society of the Pacific Conference Series, Vol. 383, Subsurface and
  Atmospheric Influences on Solar Activity, ed. {R.~Howe, R.~W.~Komm,
  K.~S.~Balasubramaniam, \& G.~J.~D.~Petrie }, 221--226

\bibitem[{{Hurford} {et~al.}(2002)}]{hurford02}
{Hurford}, G.~J., {et~al.} 2002, \solphys, 210, 61

\bibitem[{{Jing} {et~al.}(2008){Jing}, {Wiegelmann}, {Suematsu}, {Kubo}, \&
  {Wang}}]{jing08}
{Jing}, J., {Wiegelmann}, T., {Suematsu}, Y., {Kubo}, M., \& {Wang}, H. 2008,
  \apjl, 676, L81

\bibitem[{{Leka} {et~al.}(2009){Leka}, {Barnes}, \& {Crouch}}]{leka09b}
{Leka}, K.~D., {Barnes}, G., \& {Crouch}, A. 2009, in Astronomical Society of
  the Pacific Conference Series, Vol. 415, The Second Hinode Science Meeting:
  Beyond Discovery-Toward Understanding, ed. {B.~Lites, M.~Cheung, T.~Magara,
  J.~Mariska, \& K.~Reeves}, 365--+

\bibitem[{{Leka} {et~al.}(1993){Leka}, {Canfield}, {McClymont}, {de La
  Beaujardiere}, {Fan}, \& {Tang}}]{leka93}
{Leka}, K.~D., {Canfield}, R.~C., {McClymont}, A.~N., {de La Beaujardiere},
  J.-F., {Fan}, Y., \& {Tang}, F. 1993, \apj, 411, 370

\bibitem[{{Lemen} {et~al.}(2011)}]{leman11}
{Lemen}, J.~R., {et~al.} 2011, \solphys, 172

\bibitem[{{Li} {et~al.}(2011){Li}, {Jing}, {Fan}, \& {Wang}}]{li10}
{Li}, Y., {Jing}, J., {Fan}, Y., \& {Wang}, H. 2011, \apjl, 727, L19

\bibitem[{{Li} {et~al.}(2009){Li}, {Jing}, {Tan}, \& {Wang}}]{li09}
{Li}, Y., {Jing}, J., {Tan}, C., \& {Wang}, H. 2009, Sci. China G, 52, 1702

\bibitem[{{Lin} {et~al.}(2002)}]{lin02}
{Lin}, R.~P., {et~al.} 2002, \solphys, 210, 3

\bibitem[{{Liu} {et~al.}(2005){Liu}, {Deng}, {Liu}, {Falconer}, {Goode},
  {Denker}, \& {Wang}}]{liu05}
{Liu}, C., {Deng}, N., {Liu}, Y., {Falconer}, D., {Goode}, P.~R., {Denker}, C.,
  \& {Wang}, H. 2005, \apj, 622, 722

\bibitem[{{Liu} {et~al.}(2010{\natexlab{a}}){Liu}, {Lee}, {Jing}, {Liu},
  {Deng}, \& {Wang}}]{liu10}
{Liu}, C., {Lee}, J., {Jing}, J., {Liu}, R., {Deng}, N., \& {Wang}, H.
  2010{\natexlab{a}}, \apjl, 721, L193

\bibitem[{{Liu} {et~al.}(2010{\natexlab{b}}){Liu}, {Liu}, {Wang}, {Deng}, \&
  {Wang}}]{liur10}
{Liu}, R., {Liu}, C., {Wang}, S., {Deng}, N., \& {Wang}, H. 2010{\natexlab{b}},
  \apjl, 725, L84

\bibitem[{{Liu} \& {Wang}(2009)}]{liur+wang09}
{Liu}, R., \& {Wang}, H. 2009, \apjl, 703, L23

\bibitem[{{Liu} \& {Wang}(2010)}]{liur+wang10}
---. 2010, \apjl, 714, L41

\bibitem[{{Liu} {et~al.}(2009){Liu}, {Wang}, \& {Alexander}}]{liur+implosion09}
{Liu}, R., {Wang}, H., \& {Alexander}, D. 2009, \apj, 696, 121

\bibitem[{{Liu} {et~al.}(2011){Liu}, {Hoeksema}, {Hayashi}, {Sun}, {Schuck}, \&
  {Muglach}}]{liuy11}
{Liu}, Y., {Hoeksema}, J., {Hayashi}, K., {Sun}, X., {Schuck}, P., \&
  {Muglach}, K. 2011, in AAS/Solar Physics Division Abstracts \#42, 2102--+

\bibitem[{{Melrose}(1997)}]{melrose97}
{Melrose}, D.~B. 1997, \apj, 486, 521

\bibitem[{{Metcalf} {et~al.}(2008){Metcalf}, {De Rosa}, {Schrijver}, {Barnes},
  {van Ballegooijen}, {Wiegelmann}, {Wheatland}, {Valori}, \&
  {McTtiernan}}]{metcalf08}
{Metcalf}, T.~R., {De Rosa}, M.~L., {Schrijver}, C.~J., {Barnes}, G., {van
  Ballegooijen}, A.~A., {Wiegelmann}, T., {Wheatland}, M.~S., {Valori}, G., \&
  {McTtiernan}, J.~M. 2008, \solphys, 247, 269

\bibitem[{{Metcalf} {et~al.}(2006)}]{metcalf06}
{Metcalf}, T.~R., {et~al.} 2006, \solphys, 237, 267

\bibitem[{{Moore} {et~al.}(1995){Moore}, {Larosa}, \& {Orwig}}]{moore95}
{Moore}, R.~L., {Larosa}, T.~N., \& {Orwig}, L.~E. 1995, \apj, 438, 985

\bibitem[{{Moore} {et~al.}(2001){Moore}, {Sterling}, {Hudson}, \&
  {Lemen}}]{moore01}
{Moore}, R.~L., {Sterling}, A.~C., {Hudson}, H.~S., \& {Lemen}, J.~R. 2001,
  \apj, 552, 833

\bibitem[{{Patterson} \& {Zirin}(1981)}]{patterson81}
{Patterson}, A., \& {Zirin}, H. 1981, \apjl, 243, L99

\bibitem[{{Priest} \& {Forbes}(2002)}]{priest02}
{Priest}, E.~R., \& {Forbes}, T.~G. 2002, \aapr, 10, 313

\bibitem[{{Qiu} \& {Gary}(2003)}]{qiu03}
{Qiu}, J., \& {Gary}, D.~E. 2003, \apj, 599, 615

\bibitem[{{Schmieder} {et~al.}(1994){Schmieder}, {Hagyard}, {Guoxiang},
  {Hongqi}, {Kalman}, {Gyori}, {Rompolt}, {Demoulin}, \&
  {Machado}}]{schmieder94}
{Schmieder}, B., {Hagyard}, M.~J., {Guoxiang}, A., {Hongqi}, Z., {Kalman}, B.,
  {Gyori}, L., {Rompolt}, B., {Demoulin}, P., \& {Machado}, M.~E. 1994,
  \solphys, 150, 199

\bibitem[Schou et al.(2011)]{schou11} Schou, J., Scherrer, 
P.~H., Bush, R.~I., et al.\ 2011, \solphys, tmp, 368


\bibitem[{{Schuck}(2008)}]{schuck08}
{Schuck}, P.~W. 2008, \apj, 683, 1134

\bibitem[{{Sun} {et~al.}(2011){Sun}, {Hoeksema}, {Liu}, {Wiegelmann}, \&
  {Hayashi}}]{sun11}
{Sun}, X., {Hoeksema}, T., {Liu}, Y., {Wiegelmann}, T., \& {Hayashi}, K. 2011,
  in AAS/Solar Physics Division Abstracts \#42, 2101--+

\bibitem[{{Tsuneta} {et~al.}(2008)}]{tsuneta08}
{Tsuneta}, S., {et~al.} 2008, \solphys, 249, 167

\bibitem[{{Wang}(1992)}]{wang92}
{Wang}, H. 1992, \solphys, 140, 85

\bibitem[{{Wang}(2006)}]{wang06}
---. 2006, \apj, 649, 490

\bibitem[{{Wang} {et~al.}(1994){Wang}, {Ewell}, {Zirin}, \& {Ai}}]{wang94}
{Wang}, H., {Ewell}, Jr., M.~W., {Zirin}, H., \& {Ai}, G. 1994, \apj, 424, 436

\bibitem[{{Wang} \& {Liu}(2010)}]{wang10}
{Wang}, H., \& {Liu}, C. 2010, \apjl, 716, L195

\bibitem[{{Wang} {et~al.}(2005){Wang}, {Liu}, {Deng}, \& {Zhang}}]{wang+liu05}
{Wang}, H., {Liu}, C., {Deng}, Y., \& {Zhang}, H. 2005, \apj, 627, 1031

\bibitem[{{Wang} {et~al.}(2007){Wang}, {Liu}, {Jing}, \&
  {Yurchyshyn}}]{wang07a}
{Wang}, H., {Liu}, C., {Jing}, J., \& {Yurchyshyn}, V. 2007, \apj, 671, 973

\bibitem[{{Wang} {et~al.}(2004{\natexlab{a}}){Wang}, {Liu}, {Qiu}, {Deng},
  {Goode}, \& {Denker}}]{wang04a}
{Wang}, H., {Liu}, C., {Qiu}, J., {Deng}, N., {Goode}, P.~R., \& {Denker}, C.
  2004{\natexlab{a}}, \apjl, 601, L195

\bibitem[{{Wang} {et~al.}(2004{\natexlab{b}}){Wang}, {Qiu}, {Jing}, {Spirock},
  {Yurchyshyn}, {Abramenko}, {Ji}, \& {Goode}}]{wang04}
{Wang}, H., {Qiu}, J., {Jing}, J., {Spirock}, T.~J., {Yurchyshyn}, V.,
  {Abramenko}, V., {Ji}, H., \& {Goode}, P.~R. 2004{\natexlab{b}}, \apj, 605,
  931

\bibitem[{{Wang} {et~al.}(2002){Wang}, {Spirock}, {Qiu}, {Ji}, {Yurchyshyn},
  {Moon}, {Denker}, \& {Goode}}]{wang02b}
{Wang}, H., {Spirock}, T.~J., {Qiu}, J., {Ji}, H., {Yurchyshyn}, V., {Moon},
  Y., {Denker}, C., \& {Goode}, P.~R. 2002, \apj, 576, 497

\bibitem[{{Wang} {et~al.}(2006){Wang}, {Song}, {Jing}, {Yurchyshyn}, {Deng},
  {Zhang}, {Falconer}, \& {Li}}]{wang06shear}
{Wang}, H.-M., {Song}, H., {Jing}, J., {Yurchyshyn}, V., {Deng}, Y.-Y.,
  {Zhang}, H.-Q., {Falconer}, D., \& {Li}, J. 2006, Chin. J. of Astron.
  Astrophys., 6, 477

\bibitem[{{Wang} {et~al.}(2011){Wang}, {Liu}, {Liu}, {Deng}, {Liu}, \&
  {Wang}}]{wang+shuo11}
{Wang}, S., {Liu}, C., {Liu}, R., {Deng}, N., {Liu}, Y., \& {Wang}, H. 2011, ApJL, submitted

\bibitem[{{Wiegelmann}(2004)}]{wiegelmann04}
{Wiegelmann}, T. 2004, \solphys, 219, 87

\bibitem[{{Wiegelmann} {et~al.}(2006){Wiegelmann}, {Inhester}, \&
  {Sakurai}}]{wiegelmann06}
{Wiegelmann}, T., {Inhester}, B., \& {Sakurai}, T. 2006, \solphys, 233, 215

\bibitem[{{Xu} {et~al.}(2010){Xu}, {Jing}, {Cao}, \& {Wang}}]{xu10}
{Xu}, Y., {Jing}, J., {Cao}, W., \& {Wang}, H. 2010, \apjl, 709, L142

\end{thebibliography}
\end{document}